%% file: main.tex
\documentclass[twocolumn,showpacs,aps,prd,floatfix]{revtex4}
\usepackage{graphicx}
\usepackage{xspace}
\usepackage{amsmath}
\usepackage{verbatim}

\input babarsym

\input shortcuts
\input numbers

\long\def\inst#1{\par\nobreak\kern 4pt\nobreak
    {\it #1}\par\vskip 10pt plus 3pt minus 3pt}

\newcommand{\BaBarYear}       {10}
\newcommand{\BaBarNumber}     {014}
\newcommand{\SLACPubNumber} {14307}

\begin{document}

\par\vskip 7cm


\noindent
\babar-PUB-\BaBarYear/\BaBarNumber \\
SLAC-PUB-\SLACPubNumber \\

\title{\boldmath Studies of \etaK and \etaPi at \babar\ and a search for a 
second-class current}

\input{pubboard/authors_jun2010_bad2048.tex}

\begin{abstract}

\pacs{11.30.Ly, 13.35.Dx, 14.60.Fg}We report on analyses of tau lepton decays
\etaK and \etaPi, with \etaDecay,
using \Nlumi \fbarn of data from the \babar\ experiment at PEP-II,
collected at center-of-mass energies at and near the $\Upsilon(4S)$ 
resonance. We measure the branching fraction for the \etaK
decay mode, $\Br(\etaK) = \NBFetaKFinal$, and report a 
95\% confidence level upper limit for 
the second-class current process \etaPi, $\Br(\etaPi) < \NLetaPi$.
\end{abstract}

\maketitle

\input{paper}

\section{Acknowledgments}
\input{acknowledgements.tex}

\end{document}

%% file: shortcuts.tex
\def\Dm     {\ensuremath{D^-}\xspace}

\def\Dstarm {\ensuremath{D^{*-}}\xspace}

\providecommand{\Bz}{\ensuremath{B^0}\xspace}
\providecommand{\Bzb}{\ensuremath{\bar{B^0}}\xspace}

\providecommand{\Dz}{\ensuremath{D^0}\xspace}
\providecommand{\Dp}{\ensuremath{D^+}\xspace}
\providecommand{\Dm}{\ensuremath{D^-}\xspace}

\providecommand{\piz}{\ensuremath{\pi^0}\xspace}
\providecommand{\pip}{\ensuremath{\pi^+}\xspace}
\providecommand{\pim}{\ensuremath{\pi^-}\xspace}

\providecommand{\Br}{\ensuremath{\mathcal{B}}\xspace}

\providecommand{\epem}  {\ensuremath{e^+e^-}\xspace}

\providecommand{\uds}   {\ensuremath{uds}\xspace}

\providecommand{\et}   {\ensuremath{\eta}\xspace}
\providecommand{\ta}   {\ensuremath{\tau}\xspace}
\providecommand{\tp}   {\ensuremath{\tau^+}\xspace}
\providecommand{\tm}   {\ensuremath{\tau^-}\xspace}

\providecommand{\tptm} {\ensuremath{\tau^+\tau^-}\xspace}

\providecommand{\nut}  {\ensuremath{\nu_{\tau}}\xspace}
\providecommand{\qqbar}{\ensuremath{q\bar{q}}\xspace}

\providecommand{\etag}  {\ensuremath{e}-tag\xspace}

\providecommand{\mtag}  {\ensuremath{\mu}-tag\xspace}
\providecommand{\mtagged}  {\ensuremath{\mu}-tagged\xspace}

\providecommand{\fbarn}{\ensuremath{\mbox{fb}^{-1}}\xspace}

\providecommand{\GeVcc}{\ensuremath{\mbox{GeV/}c^2}\xspace}

\providecommand{\MeVcc}{\ensuremath{\mbox{MeV/}c^2}\xspace}

\providecommand{\threePi}{\ensuremath{\pip\pim\piz}\xspace}
\providecommand{\fourPi}{\ensuremath{\pip\pim\piz\pim}\xspace}
\providecommand{\threePiK}{\ensuremath{\pip\pim\piz\Km}\xspace}

\providecommand{\fourPiMode}{\ensuremath{\tm\to\pip\pim\piz\pim\nut}\xspace}

\providecommand{\threePiKMode}{\ensuremath{\tm\to\pip\pim\piz\Km\nut}\xspace}

\providecommand{\etaDecay}{\ensuremath{\eta\to\pip\pim\piz}\xspace}

\providecommand{\etaPi}{\ensuremath{\tm\to\eta\pim\nut}\xspace}
\providecommand{\etaK}{\ensuremath{\tm\to\eta\Km\nut}\xspace}
\providecommand{\etaPiPiz}{\ensuremath{\tm\to\eta\pim\piz\nut}}
\providecommand{\etaPiKz}{\ensuremath{\tm\to\eta K^{0}\pim\nut}}
\providecommand{\etaKPiz}{\ensuremath{\tm\to\eta\Km\piz\nut}}

%% file: numbers.tex
\providecommand{\Nlumi}{\ensuremath{ 470 }\xspace}

\providecommand{\NBFetaKsys}{\ensuremath{(1.48\pm 0.15\pm 0.08)\times10^{-4}}
                             \xspace}

\providecommand{\NBFetaKsysM}{\ensuremath{(1.33\pm 0.15\pm 0.09)\times10^{-4}}
                              \xspace}

\providecommand{\NBFetaKFinal}{\ensuremath{(1.42\pm0.11\text{(stat)}\pm0.07\text{(syst)})\times10^{-4}}\xspace}
\providecommand{\NBFworldAv}{\ensuremath{(1.52\pm0.08)\times10^{-4}}\xspace}

\providecommand{\NBFetaPiFinal}{\ensuremath{(3.4\pm3.4\text{(stat)}\pm2.1\text{(syst)})\times10^{-5}}\xspace}
\providecommand{\NLetaPi}{\ensuremath{9.9\times10^{-5}} \xspace}
\providecommand{\NLetaPiNinety}{\ensuremath{8.5\times10^{-5}} \xspace}

%% file: pubboard/authors_jun2010_bad2048.tex
%
\author{P.~del~Amo~Sanchez}
\author{J.~P.~Lees}
\author{V.~Poireau}
\author{E.~Prencipe}
\author{V.~Tisserand}
\affiliation{Laboratoire d'Annecy-le-Vieux de Physique des Particules (LAPP), Universit\'e de Savoie, CNRS/IN2P3,  F-74941 Annecy-Le-Vieux, France}
\author{J.~Garra~Tico}
\author{E.~Grauges}
\affiliation{Universitat de Barcelona, Facultat de Fisica, Departament ECM, E-08028 Barcelona, Spain }
\author{M.~Martinelli$^{ab}$}
\author{A.~Palano$^{ab}$ }
\author{M.~Pappagallo$^{ab}$ }
\affiliation{INFN Sezione di Bari$^{a}$; Dipartimento di Fisica, Universit\`a di Bari$^{b}$, I-70126 Bari, Italy }
% ...
\collaboration{The \babar\ Collaboration}
\noaffiliation

%% file: paper.tex
\section{INTRODUCTION}
\label{sec:intro}

Weak hadronic currents of spin-parity $J^P$ can be classified
either as first or second class according to their transformation
properties under G parity (a combination of charge conjugation and isospin
rotation)~\cite{ref:scc}. In hadronic \ta decays, the first-class
currents have $J^{PG} = 0^{++}, 0^{--}, 1^{+-}$ or $1^{-+}$ and are expected
to dominate. The second-class currents, which have
$J^{PG} = 0^{+-}, 0^{-+}, 1^{++}$ or $1^{--}$, are associated with a matrix
element proportional to the mass difference between up and down quarks. They
vanish in the limit of perfect isospin symmetry. So, while the Standard Model
does not prohibit second-class currents, such \ta decays
are expected to have branching fractions of order
$10^{-5}$~\cite{ref:pich} and no evidence has been found for them to date.

The \tm lepton provides a clean means to search for second-class currents,
through the decay mode \etaPi (charge-conjugate reactions are implied
throughout this paper).
The $\et\pim$ final state must have either $J^{PG} = 0^{+-}$ or
$J^{PG} = 1^{--}$, both of which can only be produced
via second-class currents.
The decay could be mediated by the $a_0(980)^-$ meson or by the $\pi_1(1400)^-$
resonance.
The CLEO collaboration has produced the most stringent limit so far on \etaPi
decays, finding $\Br(\etaPi) < 1.4\times 10^{-4}$
at the $95\%$ confidence level~\cite{ref:cleo}. In this work we search for the
\etaPi decay, with the \et decaying to \threePi, using the
large \ta-pair sample available from the \babar\ experiment.

The \etaK branching fraction has previously been measured by the
CLEO~\cite{ref:cleo},
ALEPH~\cite{ref:aleph} and Belle~\cite{ref:belle} Collaborations,
giving a world average value of
$\Br(\etaK) = (1.61 \pm 0.11)\times10^{-4}$~\cite{ref:pdg}.
The measurement of $\Br(\etaK)$ reported here is the first from the
\babar\ experiment, and its consistency with the Particle Data Group (PDG)
value helps to validate the method used for the \etaPi analysis.

\section{BABAR EXPERIMENT}
\label{sec:babar}

The analysis is based on data recorded by the \babar\ detector
\cite{ref:babar} at the \pep2\ asymmetric-energy \epem storage rings
operated at the SLAC National Accelerator Laboratory.
An integrated luminosity of \Nlumi\ \fbarn was collected from \epem
annihilations at and near the $\Upsilon(4S)$ resonance:
91\% of the luminosity was collected at a center-of-mass energy of
$\sqrt{s} = 10.58$~GeV, while 9\% was collected 40~MeV below this.
With a cross section of $(0.919\pm0.003)$~nb~\cite{ref:tauxsec}
for \ta-pair production at our luminosity-weighted center-of-mass
energy, the data sample contains about $432$~million produced \tptm events.

The \babar\ detector is described in detail in Ref.~\cite{ref:babar}.
Charged-particle (track) momenta are measured with a 5-layer
double-sided silicon vertex tracker (SVT) and a 40-layer drift chamber (DCH).
Outside the DCH there is a ring-imaging Cherenkov detector (DIRC) and
an electromagnetic calorimeter (EMC) consisting of 6580 CsI(Tl)
crystals. These detectors are all inside a superconducting solenoidal
magnet that produces a magnetic field of 1.5~T. Outside the magnet
there is an instrumented magnetic flux return (IFR).
In the analysis, electrons are identified from the ratio of
calorimeter energy to track momentum ($E/p$), the shape of
the shower in the calorimeter and the ionization energy
loss in the tracking system ($dE/dx$).
Muons are identified by hits in the IFR and by their small energy
deposits in the calorimeter.
Pions and kaons are identified from $dE/dx$ in the tracking
system and the Cherenkov angle from the DIRC.

\section{EVENT SELECTION}
\label{sec:selection}

Tau pairs are produced back-to-back in
the \epem center-of-mass frame, and so each event is divided into
hemispheres using the thrust axis~\cite{ref:thrust}, calculated
from all reconstructed neutral EMC clusters with an energy above 50~MeV in
the laboratory frame and all reconstructed charged particles.
Events with four well-reconstructed tracks and zero net charge
are selected. Each track is required to have a
distance of closest approach to the interaction region of
less than 10 cm when projected along
the beam axis and less than 1.5 cm in the transverse
plane. The events are required to have a ``1-3 topology'' in the
center-of-mass frame, where one track is in one hemisphere
(the tag hemisphere) and three tracks are in the other hemisphere (the
signal hemisphere). The charged particle in the tag hemisphere
must be identified as either an electron (\etag) or a muon (\mtag),
consistent with coming from a fully leptonic \ta decay.  Hadronic
tags were not used because of the large backgrounds from
$\epem\to\qqbar$ events.

The \ta candidates are reconstructed in the signal hemisphere
using the three tracks and a \piz candidate, which is reconstructed from
two separate EMC clusters, each with an energy above 30~MeV in the laboratory
frame and not associated with a charged track.
The \piz candidates are required to have an invariant mass within 15~\MeVcc
of the nominal \piz mass~\cite{ref:pdg} and are then fitted to constrain the
mass.
The \piz candidates are also required to have an energy in the laboratory
frame of at least 200~MeV.
Events with exactly one \piz candidate in the signal hemisphere, where both
EMC clusters are also in the signal hemisphere, are selected.

Backgrounds arise from a number of sources, including $\epem\to\qqbar$ events
(where $q = usdc$)
that contain \et mesons, and \ta-pair events in which a \ta
decays into a channel containing an \et meson.
The latter category includes  \etaPiPiz, \etaPiKz, \etaKPiz\ and \etaK
(background for the \etaPi mode).
These modes contribute background events when \piz or $K^0_L$ mesons
are missing, or when pions or kaons are misidentified.

To reduce backgrounds a number of other selections are applied.
The $\epem\to\qqbar$ events are suppressed by requiring the
total visible energy of the event in the lab frame to be less than
80\% of the initial-state
energy (\ta-pair events have missing energy carried by neutrinos).
This background is also suppressed by requiring the magnitude
of the event thrust in the center-of-mass frame to be greater than 0.95
(\ta-pair events at \babar\ are highly collinear).
The cosine of the angle between the thrust axis and the beam axis is
required to be less than 0.8 to ensure the selected events are well
reconstructed, without particles passing through the edges of the active
detector region near the beam pipe.
To reduce \ta background modes containing extra \piz particles
or $K^0_L$ mesons, events are rejected if they have any additional neutral
EMC clusters in the signal hemisphere with energy above 100~MeV in the
laboratory frame. After all selections, background from $b{\overline b}$
events is negligible, due mainly to the effects of the cuts on the event
multiplicity and thrust.

The overall strategy for the analysis is to fit the \threePi mass spectra
from \threePiKMode and \fourPiMode candidate events, to determine
the numbers of \etaDecay decays in the selected samples. Monte Carlo
event samples are used to estimate the numbers of \et mesons expected from
the background modes, thus allowing the contribution from the signal modes
to be determined.

The largest source of combinatorial background in the $3\pi$ mass
spectra comes from the \fourPiMode channel, which is dominated
by $\omega(782)\pim\nut$, with a significant
$\rho(770) \pi\pi\nut$ contribution. In addition, there is a small background
in the \etag sample from Bhabha events. To avoid any model dependence
in the analyses, no additional cuts are used to remove these backgrounds,
since such cuts would distort the $\eta\Km$  and $\eta\pim$ mass spectra.

\section{MONTE CARLO SIMULATIONS}
\label{sec:mc}

Monte Carlo (MC) simulations
are used to measure the signal efficiencies as well as the levels of
background. The production of \ta\ pairs is simulated with
the KK2f generator~\cite{ref:kk2f}, and  the decays of the
\ta\ lepton are modeled with Tauola~\cite{ref:tauola}.
In addition to samples of \ta-pair events in which the \ta leptons
decay according to known branching fractions,
samples of \ta\ pairs are produced for the main \ta background modes
and for the signal modes. In these dedicated samples, one \ta in each event
is decayed through the specified
mode and the other decays according to PDG branching fractions.

Continuum \qqbar\ events are separated into two samples, one for
\uubar, \ddbar and \ssbar (the \uds sample), and another for \ccbar
(the \ccbar sample). Both samples are 
simulated using JETSET~\cite{ref:jetset}, with EvtGen~\cite{ref:evtgen}
used to simulate the decays of charmed particles.
Production of $\Upsilon(4S)$ events and $B$ meson decays are
simulated using EvtGen.
Final-state radiative effects are simulated using Photos~\cite{ref:photos}.

The detector response is modeled with GEANT4~\cite{ref:geant},
and the MC events are fully reconstructed and analyzed in
the same manner as the data.

\section{ANALYSIS}
\label{sec:ana}

\subsection{The \threePi mass spectra}

In the analysis, all three charged particles in the signal
hemisphere are initially assumed to be
pions, with no requirements on the particle identification
(PID) selectors. Each event therefore has two possible \threePi
combinations. The remaining track associated with each combination
in the signal hemisphere is referred to as the `bachelor' track.

\begin{figure}[htb]
\begin{center}
\begin{tabular}{c}
\includegraphics[width=0.4\textwidth]{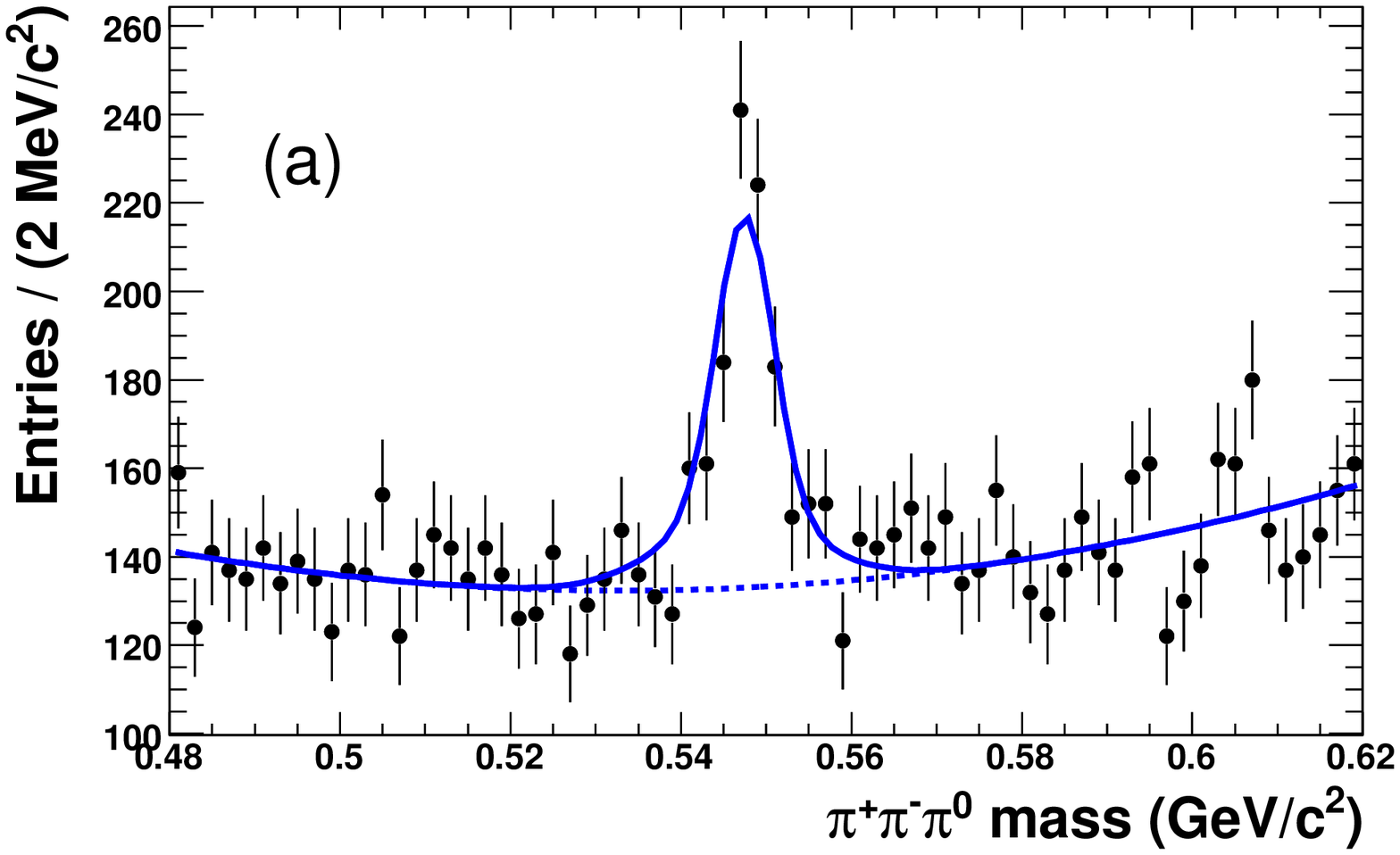} \\
\includegraphics[width=0.4\textwidth]{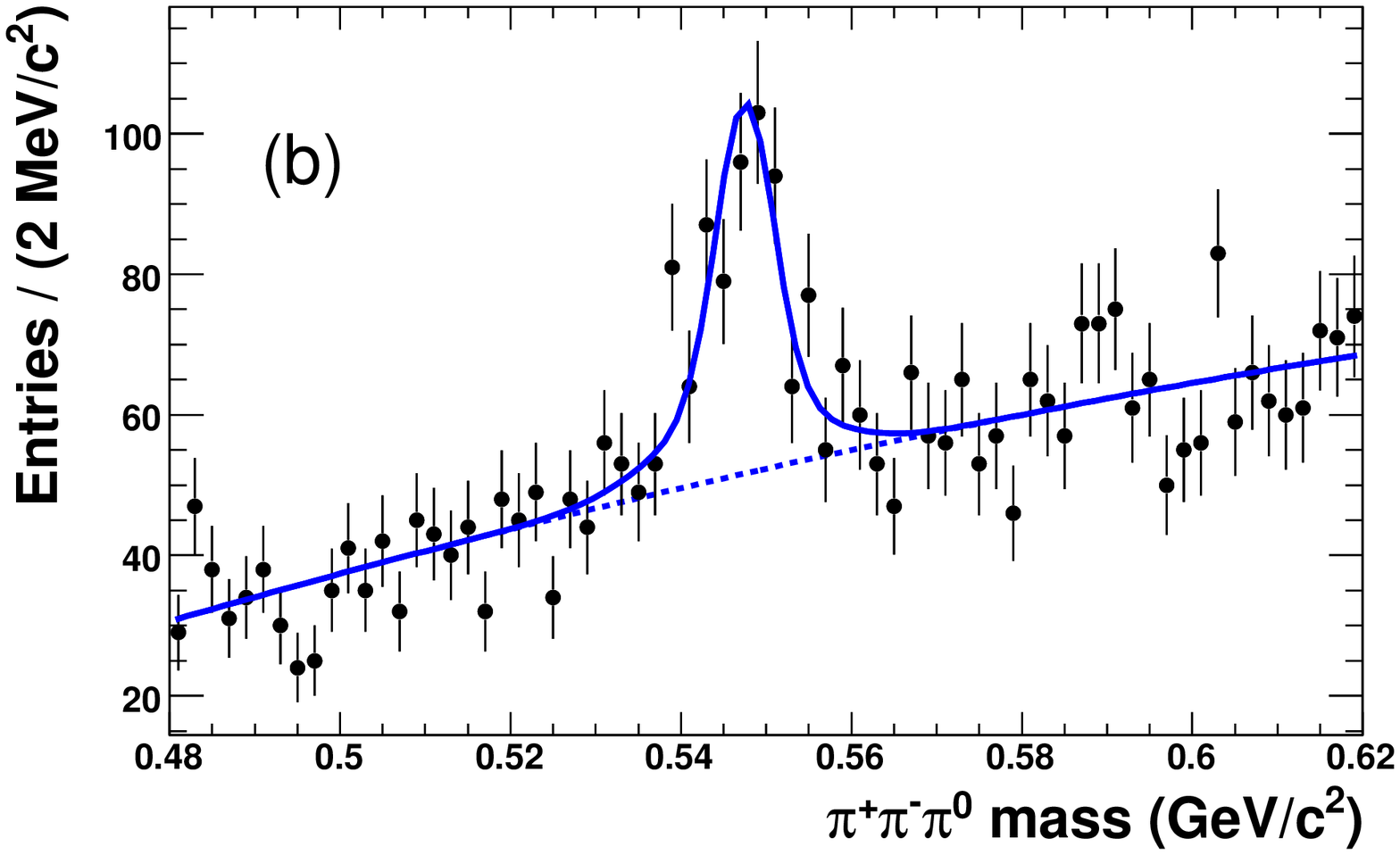}\\
\end{tabular}
\caption{Mass spectra for \threePi in \threePiKMode
candidates, for (a) \etag data and (b) \mtag data.
The curves show the results of the fits described in the text.
Note the suppressed zero on the y-axes.}
\label{fig:etaK}
\end{center}
\end{figure}

For the \etaK analysis, the bachelor track must be identified
as a kaon and the \threePiK mass is required to be less than the \ta mass.
The \threePi mass spectra with these selections are shown in
Fig.~\ref{fig:etaK} separately for the \etag and the \mtag samples;
clear $\eta$ peaks are visible in both spectra.
The curves in Fig.~\ref{fig:etaK} show the results of fits
described in Sect.~\ref{sec:fit}.

\begin{figure}[htb]
\begin{center}
\includegraphics[width=0.4\textwidth]{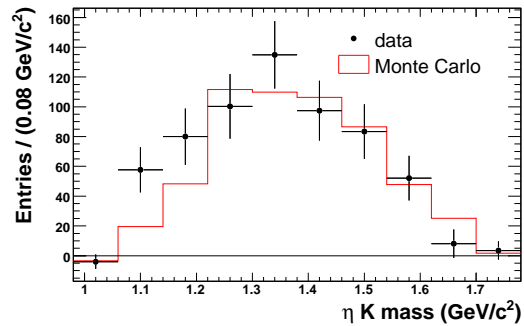}
\caption{The $\eta\Km$ mass distributions for the data and MC samples,
for $e$- and \mtag events,
obtained from the sideband subtraction method as described in the text.
The MC samples are normalized to the data luminosity; in particular the
\etaK sample is normalized to luminosity with the branching
fraction reported in this paper.}
\label{fig:etaKmass}
\end{center}
\end{figure}

The $\eta\Km$ mass distribution, as shown in Fig.~\ref{fig:etaKmass},
is constructed using a sideband subtraction method
whereby the  \threePiK mass spectrum for $3\pi$ mass in the \et sideband
regions ($0.510-0.525$ and $0.570-0.585$~\GeVcc) is  subtracted from the
spectrum where the $3\pi$ mass lies in the \et peak region
($0.54-0.555$~\GeVcc).
To correct for the shape of the combinatorial background, the entries for
the sideband
region are weighted according to factors found by intergrating over the
background functions from the fitted \threePi mass spectra.
For this figure, the various MC samples are combined according
to expected cross sections and the overall sample is normalized to the
data luminosity.
The results show agreement between data and MC, indicating that
the \etaK decay mode, which dominates the distribution, is well modeled
in Tauola.

\begin{figure}[htb]
\begin{center}
\begin{tabular}{c}
\includegraphics[width=0.4\textwidth]{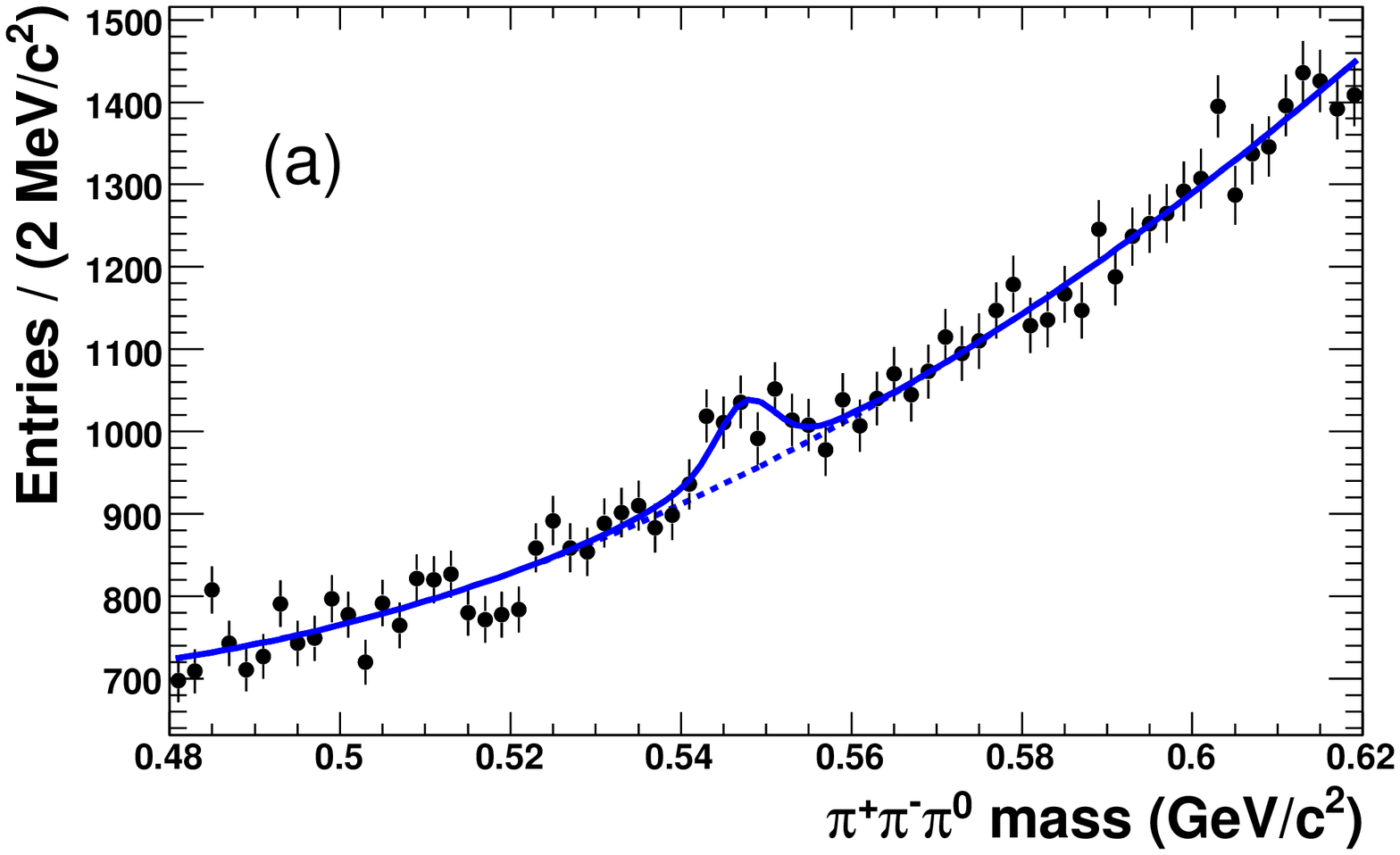} \\
\includegraphics[width=0.4\textwidth]{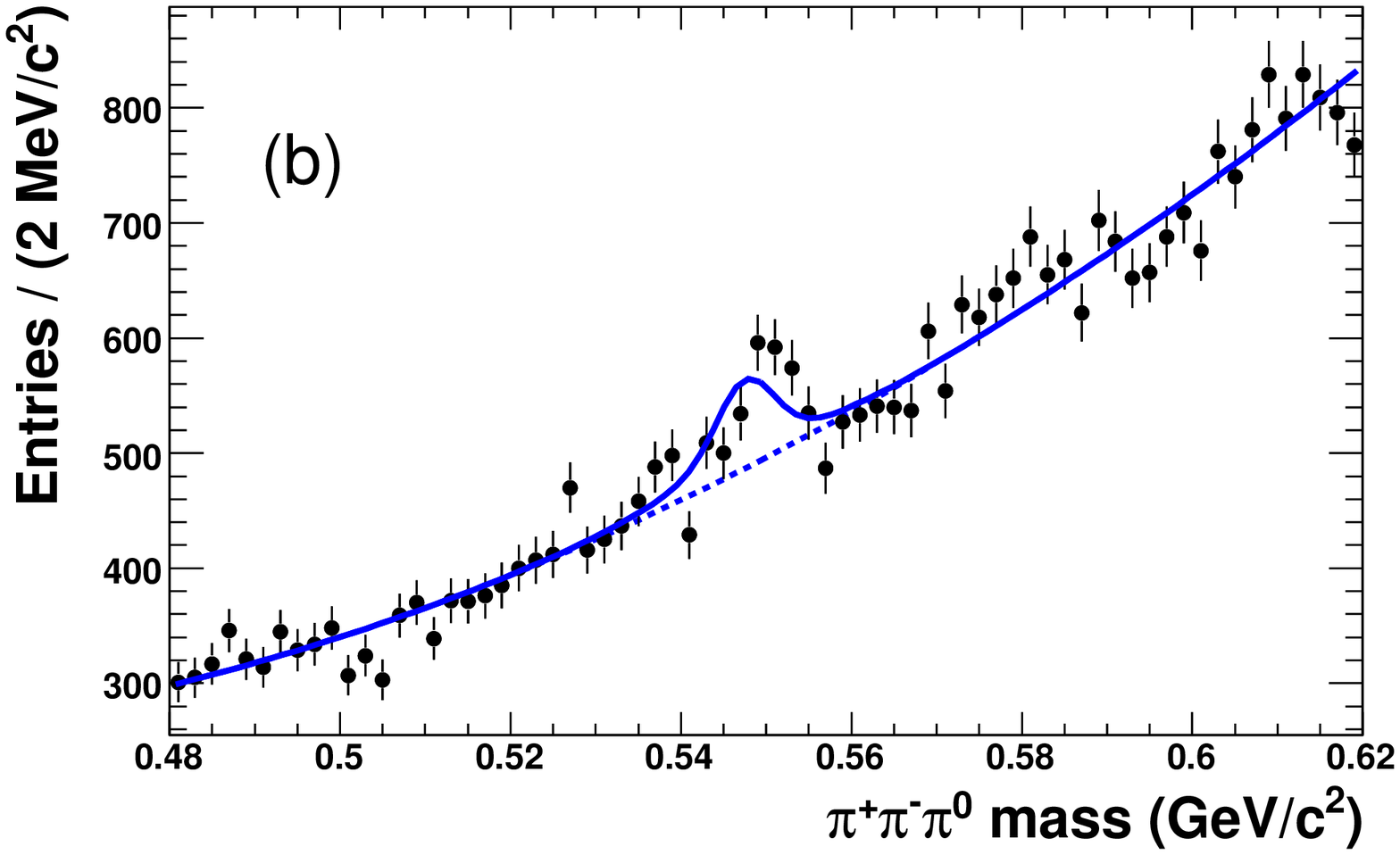}\\
\end{tabular}
\caption{Invariant \threePi mass distributions for \fourPiMode
candidates, for (a) \etag data (b) \mtag data.
The curves show the results of the fits described in the text.
Note the suppressed zero on the y-axes.}
\label{fig:etaPi}
\end{center}
\end{figure}

\begin{figure}[htb]
\begin{center}
\includegraphics[width=0.4\textwidth]{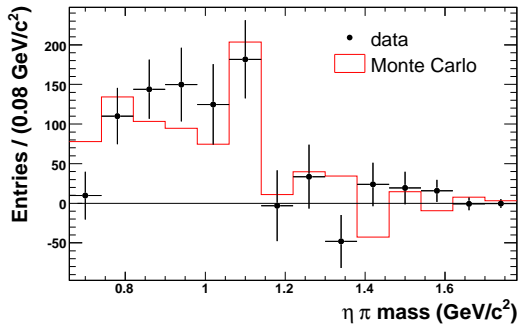}
\caption{The $\eta\pim$ mass distributions for the data and MC samples,
for $e$- and \mtag events,
obtained from the sideband subtraction method as described in the text.
The MC samples are normalized to the data luminosity; in particular,
there are no \etaPi MC events.}
\label{fig:etaPimass}
\end{center}
\end{figure}

In the search for \etaPi decays, the bachelor track must be identified
as a pion and the \fourPi mass is required to be less than the \ta mass.
The resulting \threePi mass spectra are shown in Fig.~\ref{fig:etaPi},
again separately for \etag and \mtag events. It should be noted that
while the signal \etaK channel contributes over 90\% to the \et peaks in
Fig.~\ref{fig:etaK}, the peaks in Fig.~\ref{fig:etaPi} come largely or
exclusively from backgrounds to the \etaPi search (as shown in
Tables~\ref{tab:etaK} and \ref{tab:etaPi}, to be discussed below).
Fig.~\ref{fig:etaPimass} shows the $\eta\pi$ mass distribution,
constructed using the sideband subtraction method, as described above.

\subsection{Fit parameters for the \et peaks}
\label{sec:para}

To study the shapes of the \et peaks in data and MC, high statistics
samples are examined.
The high statistics MC sample comprises the sum of $e$- and
\mtagged events from the dedicated \etaPi sample that are selected as
\fourPiMode candidates, and the $e$- and \mtagged events
from the dedicated \etaK sample that are selected as
\threePiKMode candidates.
For the data, we define a high-statistics control sample by replacing the
electron and muon tags with a charged pion tag and loosening
the selection criteria on the thrust magnitude and total event energy.
The high statistics control sample then comprises all those events that are
selected to be \fourPiMode candidates or \threePiKMode candidates.
The control sample thus defined contains a factor 20 more
\etaDecay decays than the standard data sample, coming mainly
from \uds events.

The shapes of the \et peaks in both data and MC are
found to be well described by double-Gaussian functions.
Each double-Gaussian function has five parameters: two peak masses, two
widths and a relative contribution from each Gaussian peak. The values of
these parameters are determined in fits to the high statistics samples and
are then fixed in the fits to the signal-candidate
data (Figs.~\ref{fig:etaK} and \ref{fig:etaPi}) and MC samples.
For the data sample, the core Gaussian has a width of $(3.4\pm0.1)$~\MeVcc
and a relative contribution of $62\pm4 \%$. For the MC sample, the core
Gaussian has a width of $(3.8\pm0.1)$~\MeVcc and a relative contribution
of $71\pm2 \%$.

\subsection{Fits to the mass spectra}
\label{sec:fit}

To measure the number of \et mesons in the data and MC samples, the \threePi
mass spectra are fitted over the range 0.48~\GeVcc to 0.62~\GeVcc using a
binned maximum likelihood fit.
The background is modelled as a second-order polynomial while
the \et peak is modelled using the double-Gaussian function.
The number of events in the \et peak is a free parameter in the fits,
while the five parameters of the double-Gaussian function are fixed to the
values obtained by fitting to the high statistics samples, as described above.
The fit results and errors are given in Tables~\ref{tab:etaK}
and \ref{tab:etaPi}, which are discussed later in Sect.~\ref{sec:background}.

\subsection{Efficiency}
\label{sec:eff}

The efficiency to reconstruct a signal event is defined as the probability
that a genuine signal event contributes an entry to the fitted \et peak.
The \threePi mass spectra from the dedicated \etaPi and \etaK MC samples are
fitted to measure the number of reconstructed \et mesons in each sample.
The \etaK efficiency is found to be $0.336\pm0.003 \%$ for \etag and
$0.242\pm0.003 \%$ for \mtag events, giving a total efficiency of
$0.578\pm 0.004 \%$. For \etaPi the corresponding values are
$0.286\pm 0.004 \%$, $0.186\pm 0.004 \%$ and $0.472\pm 0.006 \%$.
The efficiency for the \etaK mode is higher mainly because of a
higher efficiency for the cut on the thrust magnitude.

\section{BACKGROUNDS}
\label{sec:background}

As listed in Sect.~\ref{sec:mc}, background sources of \et mesons
include  \qqbar events as well as \ta decay modes that contain \et mesons,
such as \etaPiPiz, \etaKPiz\ and \etaPiKz. To measure the branching fractions
of \etaK and \etaPi, the numbers of \et mesons obtained from the fits must 
be corrected for contributions from the background channels.

The number of \etaDecay decays contributed by each background
mode is estimated from the MC samples, as discussed further below,
and the results are summarized in Tables~\ref{tab:etaK} and \ref{tab:etaPi},
where the first errors are statistical and the second are systematic (the
systematic errors come from the uncertainties on branching fractions).

\begin{table}[tbp]
\begin{center}
\caption{
\label{tab:etaK}
The numbers of \et mesons, for $\eta\Km$ candidates,
that are expected to come from each background mode and the total
number of \et mesons seen in the data sample, as explained in
Sections~\ref{sec:background} and \ref{sec:res}.
For each entry, the first error is statistical and the second error is
systematic.}
\begin{tabular}{|c|rll|rll|}\hline
Background contribution
&\multicolumn{6}{c|}{Expected number of events}\\ \cline{2-7}
           &\multicolumn{3}{c|}{\etag}      & \multicolumn{3}{c|}{\mtag}\\
\hline
\uds       & $4.5$ & $\pm 2.7$ & $\pm 2.3$  & $8.9$ & $\pm 4.7$ & $\pm 4.5$ \\
\ccbar     & $13.8$ & $\pm 8.3$ & $\pm 3.5$ & $0.7$ & $\pm 5.5$ & $\pm 0.2$ \\
\etaPiPiz  & $13.3$ & $\pm 3.7$ & $\pm 0.7$ & $2.9$ & $\pm 2.0$ & $\pm 0.2$ \\
\etaKPiz   & $8.4$ & $\pm 0.5$& $\pm 2.1$   & $5.0$ & $\pm 0.4$ & $\pm 1.3$ \\
\etaPiKz   & $3.9$ & $\pm 0.5$ & $\pm 0.7$  & $2.3$ & $\pm 0.4$ & $\pm 0.4$ \\
\hline
Total background     & $44$& $\pm10$& $\pm5$ & $20$& $\pm8$& $\pm 5$\\
\hline
Combined $e$- and \mtag &\multicolumn{6}{c|}{$64\pm12\pm8$} \\
\hline \hline
Measured in Data
&\multicolumn{6}{c|}{Number of events in data}\\ \cline{2-7}
                     & $463$& $\pm44$& $\pm12$& $291$ & $\pm30$ & $\pm10$ \\
\hline
Combined $e$- and \mtag &\multicolumn{6}{c|}{$754\pm53\pm16$} \\
\hline \hline
Signal
&\multicolumn{6}{c|}{Measured data$-$background}\\ \cline{2-7}
                     & $419$& $\pm44$ &$\pm16$& $271$ & $\pm30$ & $\pm13$ \\
\hline
Combined $e$- and \mtag &\multicolumn{6}{c|}{$690\pm53\pm22$} \\
\hline
\end{tabular}
\end{center}
\end{table}

\begin{table}[tbp]
\begin{center}
\caption{
\label{tab:etaPi}
The numbers of \et mesons, for $\eta\pim$ candidates,
that are expected to come from each background mode and the total
number of \et mesons seen in the data sample, as explained in
Sections~\ref{sec:background} and \ref{sec:res}.
For each entry, the first error is statistical and the second error is
systematic.}
\begin{tabular}{|c|rll|rll|}\hline
Background contribution
&\multicolumn{6}{c|}{Expected number of events}\\ \cline{2-7}
           &\multicolumn{3}{c|}{\etag}     & \multicolumn{3}{c|}{\mtag}\\
\hline
\uds       & $20$ & $\pm 9$ & $\pm 14$     & $64$  & $\pm 13$ & $\pm 43$     \\
\ccbar     & $74$ & $\pm 20$ & $\pm 19$    & $54$ & $\pm 15$ & $\pm 13$      \\
\etaPiPiz  & $215$ & $\pm 14$ & $\pm 12$   & $118$ & $\pm 11$  & $\pm 7$     \\
\etaPiKz   & $100$ & $\pm 2$ & $\pm 17$    & $71$ & $\pm 2$ & $\pm 12$       \\
\etaK      & $35$ & $\pm 1$  & $\pm 2$     & $26$ & $\pm 1$ & $\pm 1$        \\
\etaKPiz   & $0.6$ & $\pm 0.2$ & $\pm 0.1$ & $0.2$ & $\pm 0.2$ & $\pm 0.1$\\
\hline
Total background & $445$ & $\pm27$  & $\pm31$    & $333$ & $\pm23$ & $\pm47$ \\
\hline
Combined $e$- and \mtag &\multicolumn{6}{c|}{$778\pm35\pm73$} \\
\hline \hline
Measured in Data
&\multicolumn{6}{c|}{Number of events in data}\\ \cline{2-7}
                    & $489$& $\pm111$ & $\pm15$ & $424$ & $\pm74$ & $\pm13$ \\
\hline
Combined $e$- and \mtag &\multicolumn{6}{c|}{$913\pm134\pm20$} \\
\hline \hline
Signal
&\multicolumn{6}{c|}{Measured data$-$background}\\ \cline{2-7}
                    & $44$& $\pm111$  & $\pm43$ & $91$ & $\pm74$ & $\pm54$ \\
\hline
Combined $e$- and \mtag &\multicolumn{6}{c|}{$135\pm134\pm83$} \\
\hline
\end{tabular}
\end{center}
\end{table}

\subsection{Background from \uds events}

Since inclusive \et production in \uds events at \babar\ energies has
not been well measured and may be poorly simulated in the JETSET
Monte Carlo,
the high-statistics data control samples, described above, are used
to correct the MC for the level of background from this source.

To correct the \uds simulation to better match the data, scaling factors are
evaluated based on ratios of the numbers of reconstructed \et mesons in
the high-statistics (\uds-enriched) data and MC samples.
The scaling factors are found to be $1.0\pm0.5$ for the
$\eta\Km$ channel and $1.5\pm1.0$ for the $\eta\pim$ channel.
The relatively large uncertainty for the scaling factor
in the $\eta\pim$ channel is a reflection of the poor simulation
of $a_0\to\eta\pim$ production in \uds events.

\subsection{Background from \ccbar events}

The simulation of \et meson production in \ccbar events is more reliable than
in \uds events, since \ccbar events always contain two charmed particles,
whose branching fractions are well known~\cite{ref:pdg}.
To calculate a \ccbar scaling factor, \fourPiMode
candidates are selected from the $e$- and $\mu$-tagged samples.
To enhance the number of \ccbar events the selection made on the thrust
magnitude is removed and events with a \fourPi mass greater than the \ta mass
are selected.

The $\eta\pim$ mass distribution is constructed using the sideband subtraction
method described above.
Peaks are observed that correspond to the $\Dm\to\eta\pim$ and
$D^-_s\to\eta\pim$ decays.
A scaling factor of  $1.2 \pm 0.3$ is found to give best agreement between
data and MC in the numbers of  \Dm and $\Dm_s$ mesons.
Although there is no evidence for  poor simulation of \et production in
\ccbar events, this is conservatively chosen as the \ccbar scaling factor 
for the \threePiK and \fourPi analyses.

\subsection{Background from \ta decays}

The numbers of \et mesons in the dedicated MC samples for each background
\ta-decay mode are calculated by fitting the \threePi mass spectra as
previously described.
These numbers, together with the numbers of events
before selections are made, the luminosities of the data and the
known branching fractions~\cite{ref:pdg}, are used to calculate the numbers
of \et mesons in the data sample that are expected to come from each
background mode.

\subsection{Uncertainties on backgrounds}

For each background mode included in Tables~\ref{tab:etaK} and \ref{tab:etaPi}
there is a statistical error, which comes from the fits to the \threePi mass
spectra arising mainly from limited MC statistics, and a systematic error
from uncertainties in branching fractions or scaling factors. When combining
the \etag and \mtag samples, correlated errors (e.g. due to branching
fraction uncertainties) are taken into account. The total statistical
and systematic errors are combined in
quadrature and propagated as systematic errors on the final measurements.

\section{RESULTS AND CONCLUSION}
\label{sec:res}

\begin{table*}[htb]
\begin{center}
\caption{\label{tab:sys}
Additional systematic uncertainties on the
\etaK and \etaPi branching fractions.}
\begin{tabular}{|c|c|c|c||c|c|c|}\hline
                                          &\multicolumn{3}{c||}{\etaK}
                                          &\multicolumn{3}{c|}{\etaPi} \\
\cline{2-7}
Source                                    &\hspace{10pt}\etag\hspace{10pt}
                                          &\hspace{10pt}\mtag\hspace{10pt}
                                          &$e$- and \mtag
                                          &\hspace{10pt}\etag\hspace{10pt}
                                          &\hspace{10pt}\mtag\hspace{10pt}
                                          &$e$- and \mtag \\
\hline
\piz efficiency                           &3.0\%&3.0\%&3.0\%&3.0\%&3.0\%&3.0\%
\\
Tracking efficiency                       &2.0\%&2.0\%&2.0\%&2.0\%&2.0\%&2.0\%
\\
\hspace{3pt}
Error on efficiency due to MC statistics
\hspace{3pt}
                                          &1.0\%&1.1\%&0.7\%&1.6\%&1.9\%&1.2\%
\\
\etag, \mtag PID                          &0.7\%&1.8\%&1.1\%&0.7\%&1.8\%&1.1\%
\\
Bachelor \Km/\pim PID                     &1.2\%&1.2\%&1.2\%&0.2\%&0.2\%&0.2\%
\\
Luminosity and $\sigma_{\tptm}$    &0.9\%&0.9\%&0.9\%&0.9\%&0.9\%&0.9\%
\\
\hline \hline
Total systematic uncertainty              &4.1\%&4.4\%&4.1\%&4.1\%&4.6\%&4.1\%
\\ \hline
\end{tabular}
\end{center}
\end{table*}

Tables~\ref{tab:etaK} and \ref{tab:etaPi} give the numbers of \et mesons
measured in data, as obtained from the fits (Sect.~\ref{sec:fit}), for
the $\eta\Km$ and $\eta\pim$ candidate samples.  The first errors
are statistical, while the second are systematic, calculated by varying the
values of the fixed parameters within their uncertainties.  In both
channels, the \etag and \mtag analyses are combined for the final phase of the
analyses.

The fits to the $\eta\Km$ data sample yield $754\pm53\pm16$ \et mesons,
compared to an expected background of $64\pm12\pm8$, giving a
signal contribution of $690\pm53\pm22$ \et mesons. For the
$\eta\pim$ sample, the fits yield $913\pm134\pm20$ \et mesons,
with an expected background of $778\pm35\pm73$, and a
signal contribution of $135\pm134\pm83$ \et mesons. The
statistical errors on the signals are taken to be the same as those on
the unsubtracted measurements, and the other error contributions
are combined to give the total systematic errors.

Additional sources of systematic uncertainties on the measurements
of branching fractions are listed in Table~\ref{tab:sys}. The uncertainty 
in the \piz detection efficiency is 3\% per \piz candidate, while
the uncertainty on the tracking efficiency for charged particles is 0.5\% per
track, which is added linearly for the four tracks.
The error on the efficiency due to MC statistics comes from the statistical
error on the fits, as given in Sect.~\ref{sec:eff}.
The uncertainties on the PID selectors are calculated from control samples to
be 0.7\% for electrons, 1.8\% for muons, 1.2\% for kaons and 0.2\% for pions.
The uncertainty on the number of \tp\tm events is 0.9\%.

The branching fraction for \etaK is measured to be
\begin{equation}
\Br(\etaK) = \NBFetaKFinal.
\end{equation}
\noindent The values obtained separately for the \etag and \mtag samples
are \NBFetaKsys  and \NBFetaKsysM.
The measurement is compatible with the world average of
$(1.61\pm 0.11)\times10^{-4}$,
which is dominated by the Belle measurement of
$(1.58 \pm 0.05 \pm 0.09)\times 10^{-4}$~\cite{ref:belle};
this Belle measurement used the $\eta\to\gamma\gamma$ and the
$\eta\to\threePi$ decay modes (a branching fraction of
$(1.60 \pm 0.15 \pm 0.10)\times 10^{-4}$ is reported from the $\et\to\threePi$
decay mode alone).
The Belle Collaboration suggest that previous \etaK
measurements~\cite{ref:cleo,ref:aleph} underestimated background
contamination, an assertion that is supported by the observation that the
Belle and \babar\ results are in good agreement. The weighted average of the
\babar\ and Belle results is
\begin{equation}
\Br(\etaK) = \NBFworldAv,
\label{equ:worldAve}
\end{equation}
where small correlations between the systematic uncertainties of the two
experiments have not been taken into account.

The branching fraction for \etaPi is measured to be
\begin{equation}
\Br(\etaPi) = \NBFetaPiFinal.
\end{equation}
\noindent With no evidence for a signal, a 95\% confidence level
upper limit is obtained using $\Br + 1.645\sigma$, where $\Br$ is the measured
\etaPi branching fraction and $\sigma$ is its total uncertainty.
We find
\begin{equation}
\Br(\etaPi) < \NLetaPi.
\end{equation}
\noindent The limit at 90\% confidence level is $\Br(\etaPi) < \NLetaPiNinety$.
This limit improves on the CLEO value~\cite{ref:cleo}, further constraining
branching fractions for second-class current processes.

%% file: acknowledgements.tex
We are grateful for the 
extraordinary contributions of our \pep2\ colleagues in
achieving the excellent luminosity and machine conditions
that have made this work possible.
The success of this project also relies critically on the 
expertise and dedication of the computing organizations that 
support \babar.
The collaborating institutions wish to thank 
SLAC for its support and the kind hospitality extended to them. 
This work is supported by the
US Department of Energy
and National Science Foundation, the
Natural Sciences and Engineering Research Council (Canada),
the Commissariat \`a l'Energie Atomique and
Institut National de Physique Nucl\'eaire et de Physique des Particules
(France), the
Bundesministerium f\"ur Bildung und Forschung and
Deutsche Forschungsgemeinschaft
(Germany), the
Istituto Nazionale di Fisica Nucleare (Italy),
the Foundation for Fundamental Research on Matter (The Netherlands),
the Research Council of Norway, the
Ministry of Education and Science of the Russian Federation, 
Ministerio de Ciencia e Innovaci\'on (Spain), and the
Science and Technology Facilities Council (United Kingdom).
Individuals have received support from 
the Marie-Curie IEF program (European Union), the A. P. Sloan Foundation (USA) 
and the Binational Science Foundation (USA-Israel).